\documentclass[11pt,a4paper]{article}
\pdfoutput=1
\usepackage{jheppub}
\usepackage{latexsym}
\usepackage{pdflscape}
\usepackage{bm}
\numberwithin{equation}{section}
\allowdisplaybreaks

\allowdisplaybreaks[2]

\def\cO{\mathcal{O}}

\def\cW{\mathcal{W}}
\def\cB{\mathcal{B}}
\def\cS{\mathcal{S}}

\def\mint{\int_{-\infty}^\infty\!\cdots\!\int_{-\infty}^\infty}

\newcommand{\be}{\begin{equation}}
\newcommand{\ee}{\end{equation}}
\newcommand{\ba}{\begin{aligned}}
\newcommand{\ea}{\end{aligned}}

\def\bra#1{\langle #1 |}
\def\ket#1{| #1 \rangle}

\def\({\left(}
\def\){\right)}

\newcommand{\pd}{\partial}

\newcommand{\re}{{\rm e}}
\newcommand{\ri}{{\rm i}}
\newcommand{\rd}{{\rm d}}

\preprint{DESY\ 15-124}

\title{Comments on Exact Quantization Conditions and Non-Perturbative Topological Strings}

\author{Yasuyuki Hatsuda}

\affiliation{DESY Theory Group, DESY Hamburg, \\
Notkestrasse 85, D-22603 Hamburg, Germany}

\emailAdd{yasuyuki.hatsuda@desy.de}

\abstract{
We give some remarks on exact quantization conditions associated with quantized mirror curves of local Calabi-Yau threefolds, 
conjectured in \texttt{arXiv:1410.3382}.
It is shown that they characterize a non-perturbative completion of the refined topological strings in the Nekrasov-Shatashvili limit.
We find that the quantization conditions enjoy an exact S-dual invariance.
We also discuss Borel summability of the semi-classical spectrum.
}

\begin{document}

\maketitle

\renewcommand{\thefootnote}{\arabic{footnote}}
\setcounter{footnote}{0}
\setcounter{section}{0}

\section{Introduction}\label{sec:intro}
String theory is defined only perturbatively. It is widely believed that non-perturbative effects in string theory 
are explained by D-branes \cite{Polchinski:1995mt}. Topological string  theory is a toy model of string theory \cite{BCOV}.
It provides us many significant insights in string theory, M-theory and gauge theories.
One natural question is how we should formulate topological string theory non-perturbatively.
There are several attempts for this goal, based on the large $N$ duality \cite{GV2},
the resurgence theory \cite{Marino2008},
a relation to supersymmetric gauge theories \cite{LV, HMMO} 
and a quantization of spectral curves \cite{GHM1}.

In \cite{GHM1}, Grassi, Mari\~no and the author proposed a new perspective on the topological strings.
Using mirror symmetry, we start with a quantization of mirror curves (with genus one) for local Calabi-Yau threefolds \cite{ADKMV, ACDKV}.
These quantized mirror curves are naturally associated with trace-class operators \cite{GHM1, KaMa}.
Such operators have an infinite number of discrete eigenvalues.
In \cite{GHM1}, the exact spectral determinants for these trace-class operators were conjectured by using the earlier results in \cite{HMMO}.
These spectral determinants solve the spectral problem, and lead to \textit{exact quantization conditions}
as a consequence.
Moreover, in this approach, the spectral determinant naturally introduces well-defined quantities, which we refer to as 
fermionic spectral traces $Z(N;\hbar)$.
As was shown in \cite{MZ, KMZ}, in some cases, these fermionic spectral traces are represented as matrix integrals.
In the 't Hooft limit: $N \to \infty$ with $N/\hbar$ fixed finite, 
the $1/N$ expansion of the ``free energy'' $\log Z(N;\hbar)$ gives the all-genus result of the (unrefined) topological
strings in the so-called conifold frame. 
The important point is that $\log Z(N;\hbar)$ also receives non-perturbative corrections at strong 't Hooft coupling 
in the $1/N$ expansion.
In this sense, the fermionic spectral trace $Z(N;\hbar)$ provides us a non-perturbative realization of the (unrefined) 
topological strings.
The conjecture in \cite{GHM1} was confirmed for many examples \cite{KaMa, Hatsuda, MZ, KMZ, GKMR} 
and generalized for higher genus mirror curves \cite{CGM}. 

In this note, we see another aspect of \cite{GHM1}.
We focus on the exact quantization condition.
As was shown in \cite{WZH}, the exact quantization condition in \cite{GHM1} is simply written as
\be
\frac{\pd}{\pd t}\cW(t, \bm{m} ;\hbar)=2\pi\(n+\frac{1}{2}\),\qquad n=0,1,2,\dots,
\label{eq:eQC}
\ee
where $t$ and $\bm{m}$ are moduli of the Calabi-Yau manifold.
As explained in \cite{HKP, HKRS}, for genus one mirror curves, there is only the single ``true'' modulus $t$, and
the others $\bm{m}$ can be regarded as ``mass'' parameters. 
The claim in \cite{GHM1} is that the quantization condition \eqref{eq:eQC} determines all the eigenvalues of
the trace-class operators of genus one mirror curves for arbitrary $\hbar$.
This is a quite surprising result from the viewpoint of spectral theory.
What is the meaning of the function $\cW(t, \bm{m};\hbar)$?
As discussed in \cite{ACDKV}, in the semi-classical limit $\hbar \to 0$, $\cW(t, \bm{m};\hbar)$ is
determined by the refined topological string free energy in the so-called Nekrasov-Shatashvili (NS) limit \cite{NS1}.
Note that the exact quantization condition \eqref{eq:eQC} receives not only the perturbative semi-classical corrections
but also non-perturbative corrections in the M-theoretic limit: $t \to \infty$ with fixed $\hbar$.

The main claim in this note is that the function $\cW(t, \bm{m};\hbar)$ is interpreted as
a \textit{non-perturbative free energy} of the refined topological strings in the NS limit.%
\footnote{The similar spirit is found in the context of 4d gauge theories \cite{Krefl1, Krefl2, BD, KPT}.
We note that the non-perturbative structure here looks quite different from the one there.
}
We will show that $\cW(t, \bm{m};\hbar)$ is indeed related to the NS limit of the proposal in \cite{LV}.
This is a natural extension of the quantization conditions
in 4d supersymmetric gauge theories \cite{NS1, MM1, MM2, Teschner} to topological strings.
We also point out that the quantization condition \eqref{eq:eQC} is exactly invariant under an S-dual transformation
\be
t \to \frac{2\pi t}{\hbar}, \qquad \bm{m} \to \frac{2\pi \bm{m}}{\hbar},\qquad  \hbar \to \frac{4\pi^2}{\hbar}.
\label{eq:S-trans}
\ee
This invariance is almost obvious by looking at the result in \cite{WZH}, and predicts a highly non-trivial
relation for the spectrum (see \eqref{eq:S-dual-rel}).
We numerically check it for some examples.
These facts suggest that the exact quantization condition \eqref{eq:eQC} characterizes another non-perturbative
aspect of the refined topological strings.

Furthermore, we address Borel summability of the semi-classical expansion of the spectrum.
We observe that the semi-classical expansion is very likely Borel summable.
We find a strong evidence that its Borel resummation reproduces the exact spectrum correctly.
This result implies that in the semi-classical analysis, the perturbative asymptotic expansion is sufficient
to reconstruct the exact answer.%
\footnote{One might think that this is inconsistent with the fact that the exact quantization condition
receives the non-perturbative correction, but there is no contradiction. 
The key point is that the semi-classical expansion around $\hbar=0$ (with fixed $t$) and the M-theoretic expansion 
around $t \to \infty$ (with fixed $\hbar$) are purely different.
The former is asymptotic, but the latter is convergent.
There is a problem on the order of the limits. 
It was observed in \cite{HO2} that the Borel resummation of the string perturbative expansion for the resolved conifold free energy
yields non-perturbative corrections if one re-expands it around the large radius point with finite string coupling.
Probably, the similar thing happens in the current case.}

The organization of this note is as follows.
In the next section, we review the proposal in \cite{GHM1}, and
write down the exact quantization condition re-expressed in \cite{WZH}.
This expression is beautiful, and one can easily see the S-dual invariance.
We also show that the function $\cW(t,\bm{m};\hbar)$ is exactly related to
the NS limit of the non-perturbative refined topological string free energy in \cite{LV}.
In section~\ref{sec:Borel}, we observe that the semi-classical spectrum is Borel summable.
Its Borel resummation shows a very good agreement with the true spectrum for finite $\hbar$.
In appendix~\ref{sec:WKB}, we propose an efficient way to compute the semi-classical spectrum
from numerics.

\section{Quantization of mirror curves and exact quantization conditions}
\subsection{Quantizing mirror curves}
Our starting point is mirror curves for local Calabi-Yau threefolds.
Throughout this note, we consider only the genus one mirror curves.
For a CY threefold $X$, the mirror curve of its mirror $\widehat{X}$ generically take the form
\be
W_X(\re^x,\re^p)=\cO_X(x,p)+\tilde{u}=0,
\ee
where $\tilde{u}$ is the ``true'' modulus of the genus one curve in the sense of \cite{HKP, HKRS}.
This mirror curve has enough information to construct the all-genus free energy of the (unrefined) topological
strings on $X$ \cite{BKMP}.
For local $\mathbb{F}_0=\mathbb{P}^1 \times \mathbb{P}^1$ and local $\mathbb{P}^2$, for instance, we have
\be
\ba
\cO_{\mathbb{F}_0}(x,p)&=\re^{x}+\re^{-x+m}+\re^p+\re^{-p},\\
\cO_{\mathbb{P}^2}(x,p)&=\re^x+\re^{p}+\re^{-x-p},
\ea
\ee
where $m$ is a mass parameter that plays the role of a complex modulus of $\mathbb{F}_0$.
For other CYs, see table 3.1 in \cite{GKMR}, for example.

Following \cite{ADKMV, ACDKV}, we want to quantize these mirror curves.
The prescription is very simple.
We replace the variables $(x,p)$ by the canonical operators $(\hat{x},\hat{p})$, which satisfy the commutation relation
\be
[\hat{x},\hat{p}]=\ri \hbar.
\ee
In this note, we consider only the case that $\hbar$ is positive real.
There is an ambiguity on ordering of the quantization.
Following \cite{GHM1}, we take Weyl's prescription:
\be
\re^{rx+sp} \to \re^{r\hat{x}+s\hat{p}}.
\ee
For other quantization procedures, additional factors of the form $\re^{\ri \alpha \hbar}$ appear,
but these factors can be absorbed by redefining the moduli parameters appropriately.
Then the quantized mirror curve is given by
\be
(\widehat{\cO}_X(\hat{x},\hat{p})+\tilde{u})\ket{\psi}=0.
\label{eq:qmirror}
\ee
where $\ket{\psi}$ is a wave function.

\subsection{Spectral problem and quantization conditions}

Now, we associate the quantized mirror curve \eqref{eq:qmirror} with the following operator
\be
\hat{\rho}_X(\hat{x},\hat{p}) := \widehat{\cO}_X(\hat{x},\hat{p})^{-1}.
\ee
As conjectured in \cite{GHM1} and shown in \cite{KaMa} for many examples, this inverse operator $\hat{\rho}_X$ is
a trace-class operator.
The operator $\hat{\rho}_X$ has an infinite number of discrete eigenvalues.
The quantum eigenvalue problem is thus given by
\be
\hat{\rho}_X \ket{\psi_n} = \lambda_n \ket{\psi_n},\qquad \lambda_0>\lambda_1>\lambda_2>\cdots.
\label{eq:QSP}
\ee
Note that the eigenvalues $\lambda_n$ are functions of $\hbar$.
Very interestingly, for some examples, $\hat{\rho}_X$ is expressed as an integral kernel in a proper representation.
In these cases, the eigenvalue problem is formulated by a Fredholm integral equation.
See \cite{KaMa, MZ, KMZ} for explicit forms of such kernels.

\paragraph{Spectral determinant.}
The main problem in this approach is to solve the spectral problem \eqref{eq:QSP}.
For this purpose, let us introduce the \textit{spectral determinant} of $\hat{\rho}_X$:
\be
\Xi_X(\kappa,\bm{m};\hbar):=\det(1+\kappa \hat{\rho}_X)=\prod_{n=0}^\infty (1+\kappa \lambda_n),
\label{eq:spec-det}
\ee
It is obvious that all the eigenvalues of $\hat{\rho}_X$ can be computed by zeros of the spectral determinant. 
The fermionic spectral trace $Z(N,\bm{m};\hbar)$ is introduced by expanding $\Xi_X(\kappa,\bm{m};\hbar)$
around $\kappa=0$,
\be
\Xi_X(\kappa,\bm{m};\hbar)=1+\sum_{N=1}^\infty \kappa^N Z(N,\bm{m};\hbar).
\ee
Note that the spectral determinant is an \textit{entire} function, and has only zeros at $\kappa=-1/\lambda_n$.

Surprisingly, as conjectured in \cite{GHM1}, we can construct the spectral determinant for any $\hbar$ by using known
topological string results!
The construction is as follows.
We start with the free energy of the refined topological strings.
Around the large radius point, the free energy is given by
\be
\ba
F^\text{ref}(\bm{t};\tau_1,\tau_2)
=\sum_{j_L,j_R}\sum_{w, d_j \geq 1}
\frac{1}{w} N_{j_L,j_R}^{\bm{d}}
\frac{\chi_{j_L}(q_L^w)\chi_{j_R}(q_R^w)}{(q_1^{w/2}-q_1^{-w/2})(q_2^{w/2}-q_2^{-w/2})}
\re^{-w \bm{d} \cdot \bm{t}},
\ea
\label{eq:F-ref}
\ee
where $N_{j_L,j_R}^{\bm{d}}$ are integers called refined BPS invariants, which are fixed in several ways, and
\be
\ba
&q_j=\re^{2\pi \ri \tau_j},\quad (j=1,2),\qquad q_L=\re^{\pi \ri (\tau_1-\tau_2)},\qquad
q_R=\re^{\pi \ri (\tau_1+\tau_2)},\\
&\chi_j(q)=\frac{q^{2j+1}-q^{-2j-1}}{q-q^{-1}} ,
\ea
\ee
We consider the following two limits:
\be
\ba
F^\text{GV}(\bm{t};g_s) &:=\lim_{\tau_2 \to -\tau} F^\text{ref}\(\bm{t};\tau=\frac{g_s}{2\pi}, \tau_2 \), \\
F^\text{NS}(\bm{t} ;\hbar)&:=\lim_{\tau_2 \to 0} (-2\pi \tau_2)F^\text{ref}\(\bm{t};\tau=\frac{\hbar}{2\pi}, \tau_2 \),
\ea
\qquad
\ee
where the former is the standard (unrefined) topological string limit, while
the latter is called the Nekrasov-Shatashvili limit.
We refer to the latter as the NS free energy.
It is well-known that the former takes the form
\be
F^\text{GV}(\bm{t};g_s)=\sum_{g \geq 0} \sum_{w, d_j \geq 1} \frac{1}{w} n^{\bm{d}}_g
\( 2\sin \frac{w g_s}{2} \)^{2g-2} \re^{-w \bm{d} \cdot \bm{t}},
\label{eq:F-GV}
\ee
where $n_g^{\bm{d}}$ are also integers called Gopakumar-Vafa invariants.
The NS free energy is given by
\be
\ba
F^\text{NS}(\bm{t} ;\hbar)
=\sum_{j_L,j_R}\sum_{w,d_j \geq1} \frac{1}{2w^2} N^{\bm{d}}_{j_L,j_R}
\frac{\sin \frac{\hbar w}{2}(2j_L+1) \sin \frac{\hbar w}{2} (2j_R+1)}{\sin^3 \frac{\hbar w}{2}} \re^{-w\bm{d}\cdot \bm{t} }.
\ea
\label{eq:F-NS}
\ee
The main ingredient to construct $\Xi_X(\kappa,\bm{m};\hbar)$ is the following function
\be
J_X(\mu,\bm{m};\hbar)=J_X^\text{WKB}(\mu, \bm{m}; \hbar)+J_X^\text{np}(\mu, \bm{m};\hbar),
\label{eq:JX}
\ee
where $\kappa=\re^\mu$ and the two terms $J_X^\text{WKB}(\mu, \bm{m}; \hbar)$ and $J_X^\text{np}(\mu, \bm{m}; \hbar)$
are related to the NS free energy \eqref{eq:F-NS} and the GV free energy \eqref{eq:F-GV}, respectively.
The subscripts ``WKB'' and ``np'' reflect the fact that the former is captured in the semi-classical analysis
 but the latter is non-perturbative in $\hbar$.
The chemical potential $\mu$ and the mass parameters $\bm{m}$ are related to the K\"ahler moduli $\bm{t}$ in a non-trivial way.
As in \cite{Marino15}, the WKB part is given by
\be
J_X^\text{WKB}(\mu, \bm{m}; \hbar)=F_X^\text{poly}(\bm{t};\hbar)+\sum_{j} \frac{t_j}{2\pi} \frac{\pd F^\text{NS}(\bm{t};\hbar)}{\pd t_j}
+\frac{\hbar^2}{2\pi} \frac{\pd}{\pd \hbar} \( \frac{F^\text{NS}(\bm{t};\hbar)}{\hbar} \),
\label{eq:JX-WKB}
\ee
where $F_X^\text{poly}(\bm{t};\hbar)$ is a cubic polynomial of $t_j$.
The non-perturbative part is written as
\be
J_X^\text{np}(\mu, \bm{m}; \hbar)=F^\text{GV} \( \frac{2\pi}{\hbar} \bm{t}+\pi \ri \bm{B};\frac{4\pi^2}{\hbar}\),
\label{eq:JX-np}
\ee
where $\bm{B}$ is a constant vector that depends on $X$.
One imprtant remark is mentioned. The WKB part \eqref{eq:JX-WKB} has an infinite number of poles for rational $\hbar/\pi$.
These poles, however, are completely cancelled by the poles of the non-perturbative part \eqref{eq:JX-np}. 
The total function \eqref{eq:JX} is always finite for arbitrary $\hbar$.
This structure was originally found in \cite{HMO2} in ABJM theory.

Now we can wirte down the conjecture in \cite{GHM1}.
By using the function $J_X(\mu,\bm{m};\hbar)$,
the spectral determinant is given by
\be
\Xi_X(\kappa,\bm{m};\hbar)=\sum_{n \in \mathbb{Z}}
\exp [ J_X(\mu+2\pi \ri n,\bm{m};\hbar) ].
\ee
This sum suggests a quantum deformation of the Jacobi theta function,
and the determinant is finally written as
\be
\Xi_X(\kappa,\bm{m};\hbar)=\re^{J_X(\mu,\bm{m};\hbar)} \Theta_X (\mu,\bm{m};\hbar).
\ee
This is the main result in \cite{GHM1}.
Though the construction in \cite{GHM1} is heuristic, it has passed many non-trivial tests \cite{KaMa, Hatsuda, MZ, KMZ, GKMR, CGM}.
The quantum deformed theta function $\Theta_X$ reduces to the Jacobi theta function when $\hbar =2\pi$.
The important consequence of this conjecture is that the vanishing condition of $\Theta_X$ 
leads to an exact quantization condition that determines all the eigenvalues $\lambda_n$.

\paragraph{Exact quantization conditions.}
It turned out in \cite{WZH} that the exact quantization condition takes a remarkably simple form.
In the following, we focus on the local $\mathbb{F}_0$ and local $\mathbb{P}^2$ as examples. 
It is straightforward to compute other examples (see \cite{GHM1, WZH}).
It was shown in \cite{KaMa, KMZ} that the inverse operators $\hat{\rho}_{\mathbb{F}_0}$ and $\hat{\rho}_{\mathbb{P}^2}$ 
can be written as integral kernels in terms of Faddeev's quantum dilogarithm. 
Here we do not use these representations. 
The exact quantization condition takes the form
\be
\Omega(E_n,\bm{m};\hbar)=\Omega^\text{WKB}(E_n,\bm{m};\hbar)+\Omega^\text{np}(E_n,\bm{m};\hbar)=2\pi\(n+\frac{1}{2}\),
\label{eq:eQC2}
\ee
where $E_n=-\log \lambda_n$.
Of course, the two functions $\Omega^\text{WKB}$ and $\Omega^\text{np}$ inherit the structure in \eqref{eq:JX}.
For the derivation of the exact quantization condition, see \cite{GHM1}.

The result in \cite{WZH} states that the two building blocks $\Omega^\text{WKB}$ and $\Omega^\text{np}$ for local $\mathbb{F}_0$ are 
written as
\be
\ba
\Omega^\text{WKB}(E,m;\hbar)&=\frac{t^2}{\hbar}-\frac{m}{\hbar}t-\frac{2\pi^2}{3\hbar}-\frac{\hbar}{6}+
f_{\mathbb{F}_0}(t,m;\hbar),\\
\Omega^\text{np}(E,m;\hbar)&=f_{\mathbb{F}_0}\( \frac{2\pi t}{\hbar},\frac{2\pi m}{\hbar};\frac{4\pi^2}{\hbar}\),
\ea
\label{eq:Omega}
\ee
where
\be
f_{\mathbb{F}_0}(t,m;\hbar)=-\sum_{j_L,j_R}\sum_{w,d_1,d_2 \geq1}\frac{d}{w}N^{d_1,d_2}_{j_L,j_R}
\frac{\sin \frac{\hbar w}{2}(2j_L+1) \sin \frac{\hbar w}{2}(2j_R+1)}{\sin^3 \frac{\hbar w}{2}} \re^{wd_2 m-wdt}
\label{eq:f-NS}
\ee
with $d=d_1+d_2$.
The K\"ahler modulus $t$ and the energy $E$ is related by the so-called \textit{quantum mirror map}:
\be
t=2E-\sum_{\ell=1}^\infty \widehat{a}_\ell(m;\hbar)\re^{-2\ell E}.
\label{eq:q-mirrormap}
\ee
The coefficients $\widehat{a}_\ell(m;\hbar)$ can be computed systematically from the quantized mirror curve as in \cite{ACDKV}.
The very first few forms are given by
\be
\ba
\widehat{a}_1(m;\hbar)&=2(1+\re^m), \\
\widehat{a}_2(m;\hbar)&=3(1+\re^{2m})+2(4+q+q^{-1})\re^{m},
\ea
\ee
where $q=\re^{\ri \hbar}$.

For local $\mathbb{P}^2$, there are no mass parameters.
The functions $\Omega^\text{WKB}(E;\hbar)$ and $\Omega^\text{np}(E;\hbar)$ are now given by
\be
\ba
\Omega^\text{WKB}(E;\hbar) &= \frac{t^2}{2\hbar}-\frac{\pi^2}{2\hbar}-\frac{\hbar}{8}+f_{\mathbb{P}^2}(t;\hbar), \\
\Omega^\text{np}(E;\hbar) &= f_{\mathbb{P}^2}\( \frac{2\pi t}{\hbar} ; \frac{4\pi^2}{\hbar} \),
\ea
\ee
where
\be
f_{\mathbb{P}^2}(t;\hbar)=-\frac{3}{2}\sum_{j_L,j_R}\sum_{w,d\geq1}(-1)^{wd} \frac{d}{w}N^{d}_{j_L,j_R}
\frac{\sin \frac{\hbar w}{2}(2j_L+1) \sin \frac{\hbar w}{2}(2j_R+1)}{\sin^3 \frac{\hbar w}{2}} \re^{-wdt}.
\label{eq:f-NS-P2}
\ee
The quantum mirror map is also given by
\be
t=3E-\sum_{\ell=1}^\infty \widehat{a}_\ell(\hbar)\re^{-3\ell E}.
\label{eq:q-mirrormap-P2}
\ee
The first few coefficients are
\be
\ba
\widehat{a}_1(\hbar)&=3(q^{1/2}+q^{-1/2}) , \qquad
\widehat{a}_2(\hbar)=3 \left[ q^2+q^{-2}+\frac{7}{2}(q+q^{-1}) +6 \right], \\
\widehat{a}_3(\hbar)&=3(q^{9/2}+q^{-9/2})+9(q^{7/2}+q^{-7/2})+36(q^{5/2}+q^{-5/2}) \\
&\quad+88(q^{3/2}+q^{-3/2})+144(q^{1/2}+q^{-1/2}).
\ea
\ee
It is easy to find that the function $f_{\mathbb{F}_0}(t,m;\hbar)$ and $f_{\mathbb{P}^2}(t;\hbar)$ are related to the NS free energy \eqref{eq:F-NS},
\be
\ba
f_{\mathbb{F}_0}(t,m;\hbar)&=2\frac{\pd}{\pd t} F^\text{NS}(t,t-m;\hbar ), \\
f_{\mathbb{P}^2}(t;\hbar)&=3\frac{\pd}{\pd t} F^\text{NS}(t+\pi \ri;\hbar ).
\ea
\ee

\subsection{S-duality}
Remarkably, the quantization condition \eqref{eq:eQC2} is exactly invariant under the S-transform \eqref{eq:S-trans}.
More explicitly, we have
\be
\Omega(t,\bm{m};\hbar)=\Omega \(\frac{2\pi t}{\hbar}, \frac{2\pi \bm{m}}{\hbar}; \frac{4\pi^2}{\hbar}\).
\ee
After this transformation, the semi-classical perturbative part and the non-perturbative part are exchanged.
As a result, there is a highly non-trivial relation between the spectra for $\hbar$ and $\hbar^\text{D}=4\pi^2/\hbar$
\be
t_n\( \frac{2\pi \bm{m}}{\hbar}; \frac{4\pi^2}{\hbar} \)=\frac{2\pi}{\hbar} t_n(\bm{m};\hbar).
\label{eq:S-dual-rel}
\ee 
Once the solution $t_n(\bm{m};\hbar)$ to the quantization condition is found, the energy spectrum $E_n(\bm{m};\hbar)$ is determined 
by the inverse of the quantum mirror map \eqref{eq:q-mirrormap} or \eqref{eq:q-mirrormap-P2}.
Therefore, we can know the spectrum for $\hbar > 2\pi$ from the one for $\hbar < 2\pi$.
The self-dual point $\hbar =2\pi$ is special.
It was observed in \cite{GHM1, KaMa} that at this point, the spectral determinant and the quantization condition
are drastically simplified.
Note that the similar S-dual structure is found in the context of vortex-antivortex factorization in 3d supersymmetric gauge theories \cite{Pasquetti}.

Let us test the relation \eqref{eq:S-dual-rel}.
We directly evaluate the numerical values of the eigenvalues, as in \cite{HW}.
We represent the operator $\widehat{\cO}_X(\hat{x},\hat{p})$ in an orthonormal basis.
A natural choice is eigenstates $\ket{\varphi_j}$ of the harmonic oscillator.
Then we obtain an infinite dimensional matrix representation
\be
(\cO_X)_{ij}=\bra{\varphi_i} \widehat{\cO}_X(\hat{x},\hat{p}) \ket{\varphi_j}.
\label{eq:matO}
\ee 
What we should do is to diagonalize it.
In practice, we truncate it to an $L \times L$ matrix.
The obtained eigenvalues must converge to the correct ones in the infinite size limit $L \to \infty$.
In this way, we can compute the numerical values of $E_n$ for finite $\hbar$. 
In tables~\ref{tab:S-dual-test1} and \ref{tab:S-dual-test2}, we show the numerical values of the spectrum in some cases.
We first evaluate the energy $E_n(\bm{m};\hbar)$ from the matrix \eqref{eq:matO}, 
and then translate it into $t_n(\bm{m};\hbar)$
by the quantum mirror map \eqref{eq:q-mirrormap} or \eqref{eq:q-mirrormap-P2}.
These tables show that the S-dual relation \eqref{eq:S-dual-rel} indeed holds.

\begin{table}[td]
\caption{A test of the S-dual relation \eqref{eq:S-dual-rel} for local $\mathbb{F}_0$.
We evaluate $E_n(m;\hbar)$ by diagonalizing the matrix \eqref{eq:matO},
then compute $t_n(m;\hbar)$ by the quantum mirror map \eqref{eq:q-mirrormap}.}
\begin{center}
\begin{tabular}{ccc||ccc}
\hline
$n$  &  $E_n(0; \sqrt{2}\pi)$  &  $E_n(0; 2\sqrt{2}\pi)$ & $n$ & $E_n(\log 3;\sqrt{2}\pi)$  &  $E_n(\sqrt{2}\log 3;2\sqrt{2}\pi)$\\
\hline 
$0$  &  $2.4605242719$  & $3.4605592910$ & $0$  &  $2.7528481020$  & $3.8720036669$   \\
$1$  &  $3.5984708773$  & $5.0869593531$ & $1$  &  $3.8838346782$  & $5.4902688202$ \\
$2$  &  $4.4628893132$  & $6.3111090796$ & $2$  &  $4.7460288539$  & $6.7114798487$ \\
$3$  &  $5.1859599369$  & $7.3339671932$ & $3$  &  $5.4678909735$  & $7.7326659160$ \\
\hline
\hline
$n$  &  $t_n(0; \sqrt{2}\pi)$  &  $\frac{1}{\sqrt{2}} t_n(0; 2\sqrt{2}\pi)$ &
$n$  &  $t_n(\log 3; \sqrt{2}\pi)$  &  $\frac{1}{\sqrt{2}} t_n(\sqrt{2}\log 3; 2\sqrt{2}\pi)$\\
\hline 
$0$  &  $4.8911716990$  & $4.8911716990$    & $0$  &  $5.4723171187$  & $5.4723171187$  \\
$1$  &  $7.1939389859$  & $7.1939389859$    & $1$  &  $7.7642746355$  & $7.7642746355$ \\
$2$  &  $8.9252467260$  & $8.9252467260$    & $2$  &  $9.4914538288$  & $9.4914538288$ \\
$3$  &  $10.3717946646$  & $10.3717946646$ & $3$  &  $10.9356394554$  & $10.9356394554$ \\
\hline
\end{tabular}
\end{center}
\label{tab:S-dual-test1}
\end{table}%

\begin{table}[td]
\caption{A test of the S-dual relation \eqref{eq:S-dual-rel} for local $\mathbb{P}^2$.}
\begin{center}
\begin{tabular}{ccc||ccc}
\hline
$n$  &  $E_n(\hbar=\sqrt{2}\pi)$  &  $E_n(\hbar=2\sqrt{2}\pi)$ & 
$n$  &  $E_n(\hbar=\sqrt{3}\pi)$  &  $E_n(\hbar=\frac{4\pi}{\sqrt{3}})$ \\
\hline 
$0$  &  $2.1766028324$  & $3.0806224107$ & $0$  &  $2.3885831779$  & $2.7592709081$ \\
$1$  &  $3.3090257612$  & $4.6797523377$ & $1$  &  $3.6489912430$  & $4.2135235449$ \\
$2$  &  $4.1416756694$  & $5.8572207716$ & $2$  &  $4.5730868762$  & $5.2805479657$ \\
$3$  &  $4.8329028200$  & $6.8347575781$ & $3$  &  $5.3395481139$  & $6.1655792983$ \\
\hline
\hline
$n$  &  $t_n(\hbar=\sqrt{2}\pi)$  &  $\frac{1}{\sqrt{2}}t_n(\hbar=2\sqrt{2})$ &
$n$  &  $t_n(\hbar=\sqrt{3}\pi)$  &  $\frac{\sqrt{3}}{2} t_n(\hbar=\frac{4\pi}{\sqrt{3}})$\\
\hline 
$0$  &  $6.5350964286$  & $6.5350964286$ & $0$  &  $7.1699619826$  & $7.1699619826$ \\
$1$  &  $9.9272547399$  & $9.9272547399$ & $1$  &  $10.947070164$  & $10.947070164$ \\
$2$  &  $12.425041606$  & $12.425041606$ & $2$  &  $13.719266658$  & $13.719266658$ \\
$3$  &  $14.498710295$  & $14.498710295$ & $3$  &  $16.018644947$  & $16.018644947$ \\
\hline
\end{tabular}
\end{center}
\label{tab:S-dual-test2}
\end{table}%

\subsection{Comparison to a non-perturbative proposal}
Let us compare the quantization condition \eqref{eq:eQC2} with \eqref{eq:Omega} to
the proposal in \cite{LV}.
The claim in \cite{LV} is that a non-perturbative completion of the refined topological string free energy is given by
\be
F^\text{ref}_\text{np}(\bm{t};\tau_1,\tau_2)=F^\text{ref}(\bm{t};\tau_1,\tau_2)-F^\text{ref} \( \frac{\bm{t}}{\tau_1};-\frac{1}{\tau_1},\frac{\tau_2}{\tau_1} \)
-F^\text{ref} \( \frac{\bm{t}}{\tau_2};-\frac{1}{\tau_2},\frac{\tau_1}{\tau_2} \).
\label{eq:F-LV}
\ee
where $F^\text{ref}(\bm{t};\tau_1,\tau_2)$ is the ``perturbative'' free energy \eqref{eq:F-ref}.

We want to take the NS limit in \eqref{eq:F-LV}.
In the limit $\tau_2 \to +0$ with $\bm{t}$ kept finite, the last term on the right hand side
in \eqref{eq:F-LV} vanishes because $\bm{t}/\tau_2 \to \infty$.
Also, since $\tau_2/\tau_1$ goes to zero in the limit $\tau_2 \to 0$ with $\tau_1$ kept finite, the second term is related to
the NS free energy.
One easily finds
\be
\ba
&\lim_{\tau_2 \to 0}
(-2\pi \tau_2) F^\text{ref} \( \frac{\bm{t}}{\tau};-\frac{1}{\tau},\frac{\tau_2}{\tau} \) \\
&=-\tau \sum_{j_L, j_R} \sum_{w, \bm{d} \geq 1} \frac{1}{2w^2}
N^{\bm{d}}_{j_L,j_R} \frac{\sin \frac{\pi w }{\tau} (2j_L+1) \sin \frac{\pi w}{\tau} (2j_R+1)}{\sin^3 \frac{\pi w}{\tau}}
\re^{-w \bm{d} \cdot \bm{t}/\tau}\\
&=\tau F^\text{NS}\( \frac{\bm{t}}{\tau};-\frac{1}{\tau}\)
\ea
\label{eq:F-NS-nppart}
\ee
Thus the non-perturbative NS free energy is given by
\be
\ba
F^\text{NS}_\text{np}(\bm{t};\tau)&:= \lim_{\tau_2 \to 0}(-2\pi \tau_2)F^\text{ref}_\text{np}(\bm{t};\tau,\tau_2) \\
&=F^\text{NS}(\bm{t};\tau)-\tau F^\text{NS}\( \frac{\bm{t}}{\tau};-\frac{1}{\tau}\).
\ea
\ee
Note that this result is consistent with the result for the resolved conifold in \cite{KrMk}.
Comparing \eqref{eq:F-NS-nppart} for local $\mathbb{F}_0$ with \eqref{eq:f-NS}, we find
\be
f_{\mathbb{F}_0}\( \frac{2\pi t}{\hbar},\frac{2\pi m}{\hbar};\frac{4\pi^2}{\hbar}\)
=-2 \tau\frac{\pd}{\pd t} F^\text{NS}\( \frac{t}{\tau},\frac{t-m}{\tau};-\frac{1}{\tau}\)\biggr|_{\tau=\frac{\hbar}{2\pi}}.
\ee
Combining all the results, we conclude that the function $\cW(t,m;\hbar)$ in \eqref{eq:eQC} for local $\mathbb{F}_0$ is precisely
related to the non-perturbative proposal of the NS free energy
\be
\cW_{\mathbb{F}_0}(t,m;\hbar)=\frac{t^3}{3\hbar}-\frac{m}{2\hbar}t^2-\( \frac{2\pi^2}{3\hbar}+\frac{\hbar}{6} \)t
+2F_\text{np}^\text{NS}(t,t-m;\hbar).
\ee
where we dropped an integration constant.
Similarly, for local $\mathbb{P}^2$, we find
\be
\cW_{\mathbb{P}^2}(t;\hbar)=\frac{t^3}{6\hbar}-\( \frac{\pi^2}{2\hbar}+\frac{\hbar}{8} \)t
+3\widetilde{F}_\text{np}^\text{NS}(t;\hbar).
\ee
where $\widetilde{F}_\text{np}^\text{NS}(t;\hbar)$ is a little bit modified non-perturbative free energy, defined by
\be
\ba
\widetilde{F}_\text{np}^\text{NS}(t;\hbar)=F^\text{NS}(t+\pi \ri;\tau)-\tau F^\text{NS}\( \frac{t}{\tau}+\pi \ri;-\frac{1}{\tau}\),
\qquad \tau = \frac{\hbar}{2\pi}
\ea
\ee
One can push the same computation for other examples.

\section{Borel summability of semi-classical spectrum}\label{sec:Borel}
In this section, we consider the WKB expansion of the spectrum, and discuss its resummation.
In the semi-classical analysis $\hbar \to 0$, the non-perturbative part in \eqref{eq:eQC2} is invisible.
The perturbative part admits the WKB expansion
\be
\Omega^\text{WKB}(E,m;\hbar)=\frac{1}{\hbar} \sum_{\ell=0}^\infty \hbar^{2\ell} \Omega_\ell(E,m).
\label{eq:Omega-WKB}
\ee
The WKB quantization condition
\be
\Omega^\text{WKB}(E_n,m;\hbar)=2\pi\(n+\frac{1}{2} \)
\label{eq:QC-WKB}
\ee
determines the semi-classical expansion of the eigenvalues
\be
\chi_n^\text{WKB}:=\re^{E_n^\text{WKB}}=\sum_{\ell=0}^\infty \hbar^\ell \chi_n^{(\ell)}.
\label{eq:chi-WKB}
\ee
In the following, we concentrate ourselves on the case of local $\mathbb{F}_0$ with $m=0$ for simplicity.
The WKB expansion up to $\hbar^3$ in this case is found in \cite{HW},
\be
\chi_n^\text{WKB}=4+(2 n+1)\hbar+\frac{2n^2+2 n+1}{8} \hbar ^2
+\frac{2n^3+3 n^2+3 n+1}{192} \hbar ^3 +\cO(\hbar^4).
\label{eq:chi-WKB2}
\ee
It is not easy to compute the higher order corrections from the quantization condition \eqref{eq:QC-WKB}.
In appendix~\ref{sec:WKB}, we propose a very efficient method to fix the coefficients $\chi_n^{(\ell)}$ for
relatively small $n$ from numerics.
Using this method, we indeed fixed $\chi_n^{(\ell)}$ ($n=0,1$) up to $\ell=36$. 
The results up to $\ell=26$ are shown in \eqref{eq:chi0-WKB} and \eqref{eq:chi1-WKB}.

\begin{figure}[tb]
\begin{center}
\begin{tabular}{cc}
\resizebox{60mm}{!}{\includegraphics{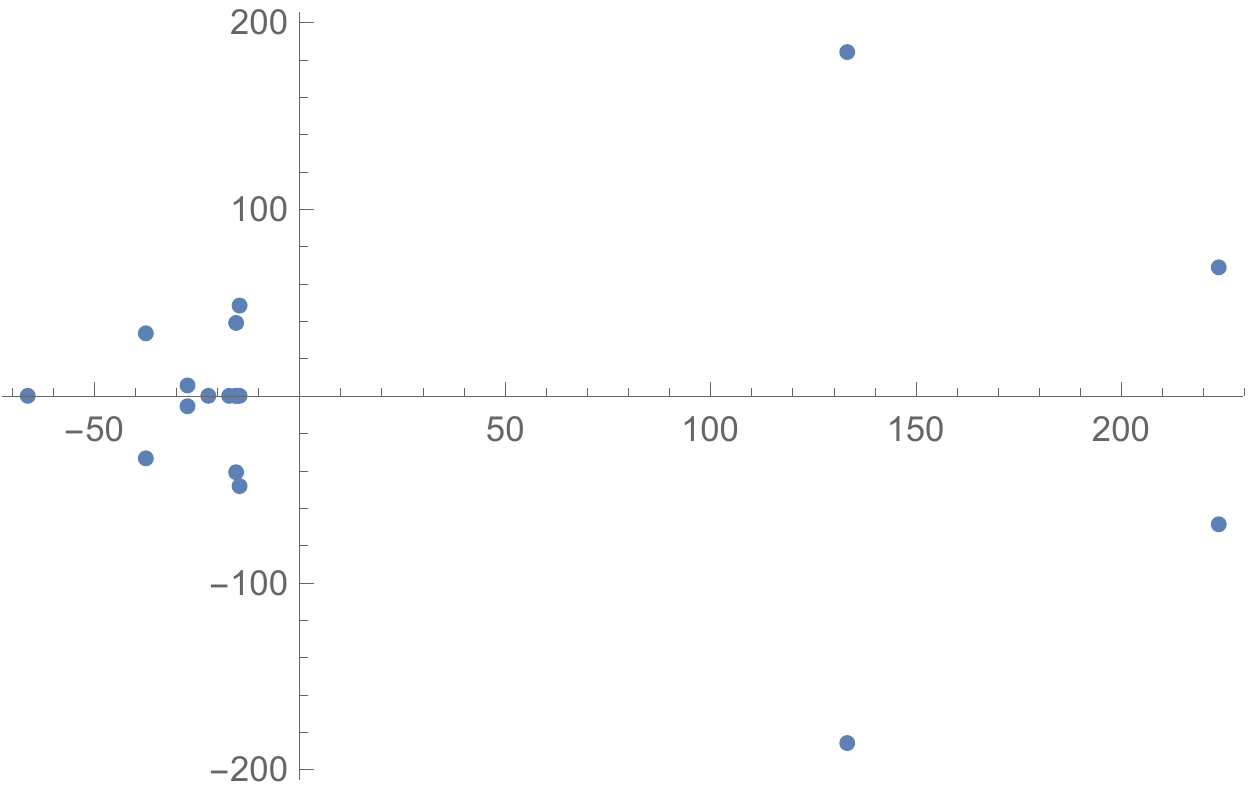}}
\hspace{10mm}
&
\resizebox{60mm}{!}{\includegraphics{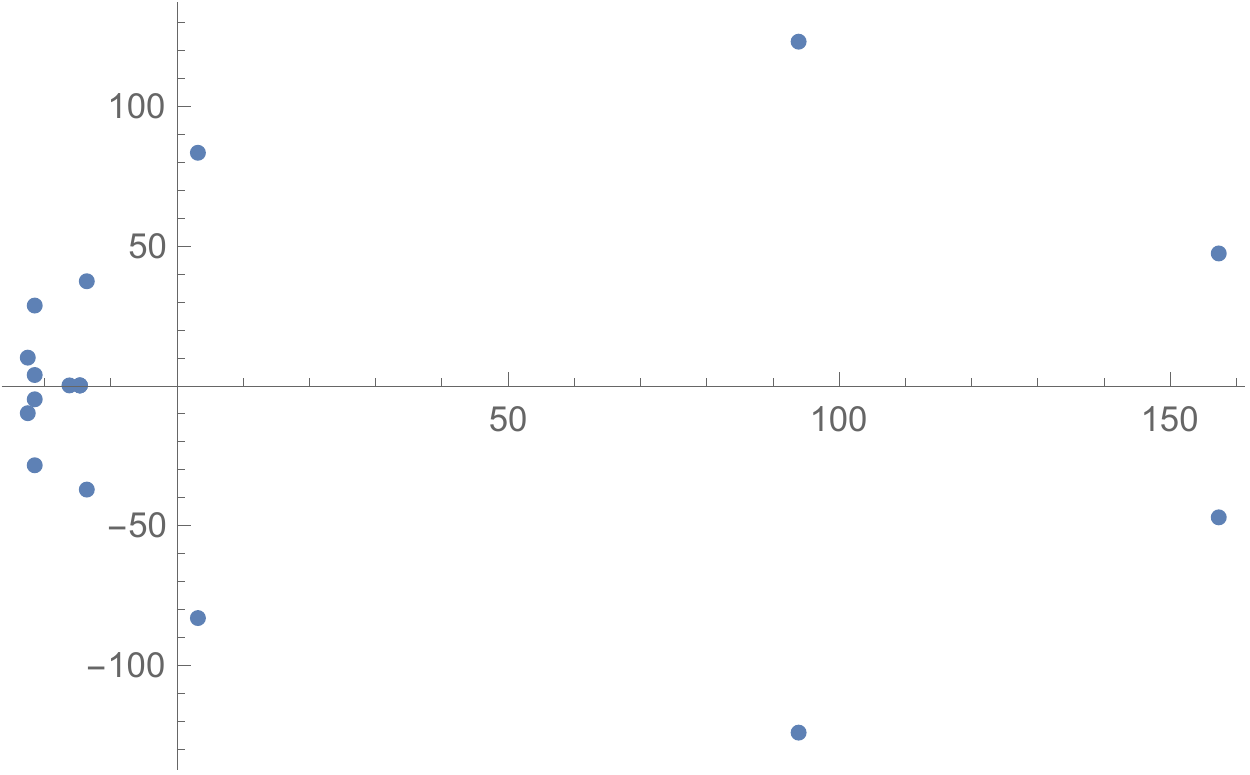}}
\\
{\footnotesize $n=0$} &
{\footnotesize $n=1$}
\end{tabular}
\end{center}
  \caption{We show the singularity structure of the Pad\'e approximant of $\cB[\chi_n^\text{WKB}](\zeta)$ ($n=0,1$).
}
  \label{fig:Borel-sing}
\end{figure}

Now, we want to perform the resummation of \eqref{eq:chi-WKB}.
We first observe that the coefficients $\chi_n^{(\ell)}$ grows factorially for $\ell \gg 1$.
Therefore, the series is asymptotic, and we need the Borel resummation.
Let us consider the Borel transform of \eqref{eq:chi-WKB}
\be
\cB [\chi_n^\text{WKB}](\zeta)=\sum_{\ell=0}^\infty \frac{\chi_n^{(\ell)}}{\ell !} \zeta^\ell.
\label{eq:Borel-trans}
\ee
The Borel resummation is then given by the inverse Laplace transform
\be
\cS_0 \chi_n^\text{WKB} = \frac{1}{\hbar} \int_0^\infty \rd \zeta \, \re^{-\zeta/\hbar} \cB [\chi_n^\text{WKB}](\zeta).
\label{eq:Borel-resum}
\ee
In the practical computation, we replace $\cB [\chi_n^\text{WKB}](\zeta)$ by its (diagonal) Pad\'e approximant,
since we have only the finite number of the coefficients $\chi_n^{(\ell)}$.
Interestingly, it is very likely that the Borel transform \eqref{eq:Borel-trans} does not have any singularities
on the positive real axis.
Therefore it is strongly expected that the series \eqref{eq:chi-WKB} is \textit{Borel summable}.
In figure~\ref{fig:Borel-sing}, we show the singularity structures of the Pad\'e approximants of $\cB [\chi_n^\text{WKB}](\zeta)$ for $n=0,1$.

Let us compare the Borel-Pad\'e resummation \eqref{eq:Borel-resum} with the numerical values of $\chi_n$ for finite $\hbar$.
In tables~\ref{tab:Borel-Pade} and \ref{tab:Borel-Pade2}, we show the results for $\hbar=\pi/2$ and $\hbar =\pi$, respectively.
From these results, we conjecture that the Borel resummation of the WKB expansion \eqref{eq:chi-WKB} reproduces
the exact values:
\be
\cS_0 \chi_n^\text{WKB} = \chi_n.
\ee
What does it imply?
This relation suggests that the Borel resummation of \eqref{eq:Omega-WKB} also coincides with the exact result
\be
\cS_0 \Omega^\text{WKB}(E,m;\hbar) \stackrel{?}{=} \Omega(E,m;\hbar).
\label{eq:guess}
\ee 
In the M-theoretic limit: $E \to \infty$ but $\hbar$ kept finite, the right hand side splits into
the two parts: the perturbative and non-perturbative parts (see \eqref{eq:Omega}).
If this guess is correct, the Borel resummation of the semi-classical expansion gives not only the perturbative resummation
in the M-theoretic expansion but also the non-perturbative correction in $\hbar$.
This structure is very similar to the resummation of the string perturbative expansion for the resolved conifold, as observed in \cite{HO2}
(see also \cite{Hatsuda}).
So far, we do not have a strong evidence of the guess \eqref{eq:guess}.
It would be very interesting to check \eqref{eq:guess}.

\begin{table}[td]
\caption{The Borel-Pad\'e resummation of $\chi_n^\text{WKB}$ ($n=0,1$) for $\hbar=\pi/2$.
The resummation using the diagonal Pad\'e approximant with order $M$ is denoted by $\cS_0^{[M/M]} \chi_n^\text{WKB}$.}
\begin{center}
\begin{tabular}{ccc}
\hline
$M$ &  $\cS_0^{[M/M]} \chi_0^\text{WKB}(\hbar=\pi/2)$ &  $\cS_0^{[M/M]} \chi_1^\text{WKB}(\hbar=\pi/2)$ \\
\hline 
$6$  &    $\underline{5.90573411}00278498180$  &  $\underline{10.48449}2226744920$ \\
$12$  &  $\underline{5.9057341149654}993842$  &  $\underline{10.48449670217}5908$ \\
$18$  &  $\underline{5.905734114965448513}1$  &  $\underline{10.4844967021727}24$ \\
\hline 
Numerical value & $5.9057341149654485137$ &     $10.484496702172730$ \\
\hline
\end{tabular}
\end{center}
\label{tab:Borel-Pade}
\end{table}%

\begin{table}[td]
\caption{The Borel-Pad\'e resummation of $\chi_n^\text{WKB}$ ($n=0,1$) for $\hbar=\pi$.}
\begin{center}
\begin{tabular}{ccc}
\hline
$M$ &  $\cS_0^{[M/M]} \chi_0^\text{WKB}(\hbar=\pi)$ &  $\cS_0^{[M/M]} \chi_1^\text{WKB}(\hbar=\pi)$ \\
\hline 
$6$  &    $\underline{8.62716}3177357$  &  $\underline{21.7}49190319$ \\
$12$  &  $\underline{8.62716891}7537$  &  $\underline{21.751895}879$ \\
$18$  &  $\underline{8.62716891392}6$  &  $\underline{21.75189571}5$ \\
\hline 
Numerical value & $8.627168913927$ &     $21.751895710$ \\
\hline
\end{tabular}
\end{center}
\label{tab:Borel-Pade2}
\end{table}%

\section{Concluding remarks}
In this note, we gave some comments on the exact quantization condition in \cite{GHM1}.
We demonstrated that it naturally provides
a non-perturbative completion of the refined topological strings in the NS limit.
Interestingly, this quantization condition enjoys the exact S-dual invariance.
This invariance gives a very strong constraint of the form of $\cW(t,\bm{m};\hbar)$.
In particular, it completely determines the non-perturbative correction from the perturbative result.
It would be interesting to check this S-dual invariance for many other examples that
has not been studied in the literature.
It is also interesting to consider what the S-duality in the exact quantization condition
implies for the spectral determinant.

In 4d supersymmetric gauge theories, the function corresponding to $\cW(t,\bm{m};\hbar)$ in \eqref{eq:eQC} 
is interpreted as the Yang-Yang potential in quantum integrable systems \cite{NS1, Teschner, NRS}.
It is interesting to explore a relation between $\cW$ here and the Yang-Yang potential.
The Nekrasov instanton partition function in the NS limit can be computed by TBA integral equations \cite{NS1} (see also \cite{MY}).
We believe that the topological string free energy in the NS limit is also governed by such integral equations.
It would be nice to find out such equations and to investigate a relation to the function $\cW$.

A generalization to higher genus mirror curves seems to be straightforward.
For a genus $g$ curve, there are $g$ ``true'' moduli $\bm{t}$.
For the genus $g$ algebraic curve, there are $g$ independent A-cycles and also $g$ independent B-cycles.
We require the single-valuedness of the wave function around every B-cycle.%
\footnote{This requirement is just an analogy with the $g=1$ case. The Bohr-Sommerfeld quantization condition
is understood as the single-valuedness of the wave function around the B-cycle in the leading WKB approximation.
The semi-classical limit of \eqref{eq:eQC-higher} should be related to the Einstein-Brillouin-Keller quantization condition.}
Then we naturally arrive at the exact quantization conditions
\be
\frac{\pd}{\pd t_i} \cW(\bm{t},\bm{m};\hbar)=2\pi \( n_i+\frac{1}{2} \), \qquad i=1,\dots,g.
\label{eq:eQC-higher}
\ee
In the semi-classical limit, the left hand side is related to the \textit{quantum B-period} for each B-cycle.
The solutions $t_i$ ($i=1,\dots,g$) to the quantization conditions \eqref{eq:eQC-higher} 
are maybe related to higher conseved charges
in the corresponding integrable system.
The quantization conditions \eqref{eq:eQC-higher} are different from the recent proposal in \cite{CGM}.
In \cite{CGM}, a generalized spectral determinant with $g$ fugacities were introduced.
As in the genus one case, it characterizes non-perturbative aspects of the topological strings through the (generalized)
fermionic spectral traces.
The consequence of the generalized spectral determinant in \cite{CGM} is that it leads to a \textit{single} quantization condition.
Up to now, it is unclear to us how these two generalizations are related to each other.
It would be nice to clarify it more deeply.

It is also interesting to consider the field theoretic limit of the topological strings.
The $A_{N-1}$ CY geometry, for example, is related to the SU($N$) Seiberg-Witten theory via the geometric engineering \cite{KKV}.
It is well-known that this theory is related to the periodic Toda chain. 
The quantization condition for the periodic Toda chain \cite{GP} 
is written in terms of the TBA equations \cite{KT}.
It would be interesting to compare the 4d field theoretic limit of \eqref{eq:eQC-higher} for the $A_{N-1}$ geometry
with the quantization condition for the Toda chain.

\acknowledgments{I thank Marcos Mari\~no, Kazumi Okuyama and J\"org Teschner for valuable discussions.
I am especially grateful to Marcos Mari\~no for helpful comments on the draft.
}

\appendix

\section{Semi-classical spectrum from numerics}\label{sec:WKB}
In this appendix, we explain how to compute the WKB expansion of the spectrum.
We know that the spectrum $\chi_n$ admits the WKB expansion \eqref{eq:chi-WKB}.
We want to fix the coefficients.
The idea here is simple.
We first compute the eigenvalues for relatively small $\hbar$ numerically
by diagonalizing the matrix \eqref{eq:matO}.
For small $\hbar$, this can be done with very high precision.
From these numerical data, one can estimate the coefficients.
To improve the numerical accuracy of the estimation, we use the Richardson extrapolation.

In the case of local $\mathbb{F}_0$ with $m=0$, we indeed computed $\chi_0$ and $\chi_1$ 
for $\hbar=1/k$ ($k=1001,1002,\dots,1100$) with 400-digit numerical precision.
Following the method in \cite{MSW}, we fixed each coefficient order by order,
and finally found the first 36 coefficients analytically.
The results up to $\hbar^{26}$ are as follows:
\be
\tiny
\ba
\chi_0^\text{WKB}&=4+\hbar+\frac{\hbar ^2}{8}+\frac{\hbar ^3}{192}+\frac{\hbar ^4}{768}-\frac{67 \hbar ^5}{245760}+\frac{653 \hbar ^6}{5898240}
-\frac{32519 \hbar^{7}}{660602880}+\frac{135001 \hbar^{8}}{5284823040}-\frac{45750727 \hbar^{9}}{3044058071040}+\frac{1198585643 \hbar^{10}}{121762322841600}\\
&\quad-\frac{151890553139 \hbar^{11}}{21430168820121600}+\frac{2860859741627 \hbar^{12}}{514324051682918400}-\frac{1011296090746987 \hbar^{13}}{213958805500094054400}+\frac{51820177352512433 \hbar^{14}}{11981693108005267046400}\\
&\quad-\frac{12196293770946645359 \hbar^{15}}{2875606345921264091136000}+\frac{102120975632292793313 \hbar^{16}}{23004850767370112729088000}-\frac{247311952643245167912847 \hbar^{17}}{50058555269797365298495488000}\\
&\quad+\frac{21003011532557603280648023 \hbar^{18}}{3604215979425410301491675136000}
-\frac{723428173784131455428496289 \hbar^{19}}{99607423431393157423042658304000} \\
&\quad+\frac{418046890256615675609799298877 \hbar^{20}}{43827266309812989266138769653760000}
-\frac{387823357100932542298961137152307 \hbar^{21}}{29451922960194328786845253207326720000} \\
&\quad+\frac{49411683325988972822114590790184413 \hbar^{22}}{2591769220497100933242382282244751360000}-\frac{2119397426588793097645787595092141123 \hbar^{23}}{73367005626379472571784359989697576960000} \\
&\quad+\frac{1046586291235684718796226704548546651251 \hbar^{24}}{22890505755430395442396720316785644011520000}-\frac{2763516923713499337845074861881834527649367 \hbar^{25}}{36624809208688632707834752506857030418432000000}\\
&\quad+\frac{493752121627987278883950960030085030222913603 \hbar ^{26}}{3808980157703617801614814260713131163516928000000}
+\cO(\hbar^{27})
\ea
\label{eq:chi0-WKB}
\ee
and
\be
\tiny
\ba
\chi_1^\text{WKB}&=4+3 \hbar+\frac{5 \hbar^{2}}{8}+\frac{3 \hbar^{3}}{64}+\frac{\hbar^{4}}{96}-\frac{257 \hbar^{5}}{81920}+\frac{2279 \hbar^{6}}{1179648}-\frac{272717 \hbar^{7}}{220200960}+\frac{4687223 \hbar^{8}}{5284823040}-\frac{78669253 \hbar^{9}}{112742891520}+\frac{2902342813 \hbar^{10}}{4870492913664}\\
&\quad-\frac{3909009230737 \hbar^{11}}{7143389606707200}+\frac{276073926226501 \hbar^{12}}{514324051682918400}-\frac{39920493269381497 \hbar^{13}}{71319601833364684800}+\frac{1481592383422891331 \hbar^{14}}{2396338621601053409280}\\
&\quad-\frac{230449325073591598519 \hbar^{15}}{319511816213473787904000}+\frac{20391142069901359300639 \hbar^{16}}{23004850767370112729088000}-\frac{19106537404213714921398917 \hbar^{17}}{16686185089932455099498496000} \\
&\quad+\frac{1118603122542169295491969877 \hbar^{18}}{720843195885082060298335027200}-\frac{804312004383307739739208157177 \hbar^{19}}{365227219248441577217823080448000} \\
&\quad+\frac{143189256891154582251747365564611 \hbar^{20}}{43827266309812989266138769653760000}-\frac{1273455094857825645294021459211783 \hbar^{21}}{251725837266618194759361138524160000} \\
&\quad+\frac{4231573452876994762093846837124778503 \hbar^{22}}{518353844099420186648476456448950272000}
-\frac{335261687140409439344796015593150731969 \hbar^{23}}{24455668542126490857261453329899192320000}\\
&\quad+\frac{2478087705885473601361764113845728412273 \hbar^{24}}{103576949119594549513107331750161285120000}-\frac{529076945753188857007972471183458908636707357 \hbar^{25}}{12208269736229544235944917502285676806144000000}\\
&\quad+\frac{61993659201098334041822970271475375738157074009 \hbar^{26}}{761796031540723560322962852142626232703385600000}
+\cO(\hbar^{27})
\ea
\label{eq:chi1-WKB}
\ee
Of course, the coefficients up to $\hbar^3$ are in agreement with \eqref{eq:chi-WKB2}.

\end{document}